\documentclass{aa}
\usepackage{txfonts}
\usepackage{natbib}
 \usepackage{times}
 \usepackage{amsfonts}
 \usepackage{amssymb}
 \usepackage{graphicx}

 \newcommand{\degree}{$^\circ$}

\newcommand{\CII}{[\ion{C}{ii}]}   
   
\newcommand{\CI}{[\ion{C}{i}]}   
   
\newcommand{\OI}{[\ion{O}{i}]}

\newcommand{\HII}{\ion{H}{ii}}

\newcommand{\lsun}{L$_{\odot}$}

\newcommand{\rasf}[4]{\ensuremath{#1^\mathrm{h}#2^\mathrm{m}#3\fs #4}}

\newcommand{\decasf}[4]{\ensuremath{#1^\circ#2^\prime#3\farcs #4}}

\begin{document}
\title{Photon Dominated Regions in NGC 3603}
\subtitle{[CI] and mid-J CO line emission}
\author{
M. R\"ollig\inst{1},
C. Kramer\inst{2},
C. Rajbahak\inst{1},
T. Minamidani\inst{3},
K. Sun\inst{1},
R. Simon\inst{1},
V. Ossenkopf\inst{1,4},
M. Cubick\inst{1},
M. Hitschfeld\inst{1},
     M.\,Aravena\inst{5} \and
     F.\,Bensch\inst{6} \and
%     A.\,Benz\inst{} \and
     F.\,Bertoldi\inst{6} \and
     L.\,Bronfman\inst{7} \and
%     M.\,Burton\inst{} \and
     M.\,Fujishita\inst{8} \and
     Y.\,Fukui\inst{9} \and
     U.U.\,Graf\inst{1} \and
     N.\,Honingh\inst{1} \and
     S.\,Ito\inst{9} \and
     H.\,Jakob\inst{10} \and
     K.\,Jacobs\inst{1} \and
     U.\,Klein\inst{6} \and
     B.-C.\,Koo\inst{11} \and
     J.\,May\inst{3} \and
     M.\,Miller\inst{1} \and
     Y.\,Miyamoto\inst{9} \and 
     N.\,Mizuno\inst{9,12} \and
     T.\,Onishi\inst{8,9} \and
     Y.-S.\,Park\inst{11} \and
     J.\,Pineda\inst{13} \and
     D.\,Rabanus\inst{14} \and
     H.\,Sasago\inst{9} \and
     R.\,Schieder\inst{1} \and
     J.\,Stutzki\inst{1} \and
     H.\,Yamamoto\inst{9} \and
     Y.\,Yonekura\inst{8}
}
\institute{I. Physikalisches Institut, Universit\"at zu K\"oln, Z\"ulpicher Str. 77, D-50937 K\"oln,
Germany
\and Instituto de Radioastronomia Milimetrica (IRAM), Avda. Divina Pastora 7, E-18012 Granada, Spain
\and Department of Physics, Faculty of Science, Hokkaido University, N10W8, Kita-ku, Sapporo 060-0810, Japan
\and SRON Netherlands Institute for Space Research, P.O. Box 800, 9700 AV 
Groningen, Netherlands
\and National Radio Astronomy Observatory. 520 Edgemont Road, Charlottesville VA 22903, USA 
\and  Argelander-Institut f\"ur Astronomie \thanks{Founded by merging of the
Sternwarte, Radioastronomisches Institut and Institut f\"ur Astrophysik und 
Extraterrestrische Forschung}, Universit\"at Bonn, Auf dem H\"ugel 71, D-53121 Bonn, Germany
\and
     Departamento de Astronom\'{i}a, Universidad de Chile, Casilla 36-D, Santiago, Chile
\and
     Department of Physical Science, Osaka Prefecture University, Gakeun 1-1, Sakai, Osaka 599-8531, Japan
\and
     Department of Astrophysics, Nagoya University, Furocho, Chikusaku, Nagoya 464-8602, Japan
\and Deutsches SOFIA Institut, Universit\"at Stuttgart, Pfaffenwaldring 31, 70569 Stuttgart, Germany        
     \and
     Seoul National University, Seoul 151-742, Korea
\and ALMA-J Project Office, National Astronomical Observatory of Japan, 2-21-1 Osawa, Mitaka, Tokyo 181-8588, Japan
\and Jet Propulsion Laboratory, M/S 169-507, 4800 Oak Grove Drive, Pasadena, CA 91109., USA
\and European Southern Observatory, Alonso de Cordova 3107, Vitacura, Casilla 19001, Santiago, Chile
}
\authorrunning{R\"ollig et al.}

\offprints{M. R\"ollig,\\ \email{roellig@ph1.uni-koeln.de}}

\abstract{}
{
  We aim at deriving the excitation conditions of the
    interstellar gas as well as the local FUV intensities in the molecular cloud surrounding NGC~3603 to get a coherent
picture of how the gas is energized by the central stars. }
{ The NANTEN2-4m submillimeter antenna is used to map the \mbox{[CI] 1-0}, \mbox{2-1} and \mbox{CO 4-3}, \mbox{7-6} lines in a $2'\times2'$ region around the young OB cluster NGC~3603~YC. These data are combined with \mbox{C$^{18}$O 2--1} data,
HIRES-processed IRAS 60 $\mu$m and 100 $\mu$m maps of the FIR continuum, and Spitzer/IRAC maps.}
{
The NANTEN2 observations show the presence of two molecular clumps located south-east and south-west of the cluster and confirm the overall structure already found by previous CS and C$^{18}$O observations.
% The two clouds also contain two pillar like structures that protrude from the molecular clouds towards the OB cluster and have already been observed in the optical and near-infrared.
We find a slight position offset of the peak intensity of CO and [CI], and the atomic carbon appears to be further extended compared to the molecular material.  
We used the HIRES far-infrared dust data to derive a map of the FUV field heating the dust. We constrain the FUV field to values of $\chi=3-6\times10^3$ in units of the Draine field across the clouds. Approximately 0.2 to 0.3 \% of the total FUV energy is re-emitted in the [CII] 158 $\mu$m cooling line observed by ISO. Applying LTE and escape probability calculations, we derive temperatures ($T_\mathrm{MM1}=43$~K, $T_\mathrm{MM2}=47$~K), column densities ($N_\mathrm{MM1}=0.9\times10^{22}$~cm$^{-2}$, $N_\mathrm{MM2}=2.5\times10^{22}$~cm$^{-2}$) and densities ($n_\mathrm{MM1}=3\times10^{3}$~cm$^{-3}$, $n_\mathrm{MM2}=10^{3}-10^{4}$~cm$^{-3}$) for the two observed molecular clumps MM1 and MM2.  
}
{The cluster is strongly interacting with the ambient molecular cloud, governing its structure and physical conditions. A stability analysis shows the existence of gravitationally collapsing gas clumps which should lead to star formation. Embedded IR sources have already been observed in the outskirts of the molecular cloud and seem to support our conclusions. }


\maketitle

\section{Introduction}
Understanding the feedback between star formation and the interstellar medium (ISM) is important in order to understand the process of star formation  \citep[see][and references therein]{zinnecker2007}. The physical and chemical properties of stars are the heritage of their parental clouds. Stars are born from interstellar gas and release metal-enriched material to the ISM when they die. Radiation from the stars is the prime heating source for gas and dust in  nearby molecular clouds and may trigger velocity and density fluctuations, that stimulate further star formation. To understand the formation of these next generation stars, it is important to understand how the previous generation interacts with its parental clouds  \citep{dale2005}. The energy incident on  the clouds in the form of stellar far ultraviolet (FUV: 6~eV$ \le h\nu \le$ 13.6~eV) radiation  is countered by cooling continuum radiation plus emission of atomic fine-structure lines (\mbox{[CII]~158~$\mu$m}, \mbox{[CI]~370~$\mu$m}, \mbox{610~$\mu$m}, and \mbox{[OI]63~$\mu$m}, \mbox{145~$\mu$m}) and by molecular rotational lines (CO, H$_2$O, OH, etc.). The cooling emission carries the imprint of the local physical and chemical conditions and can be used to infer the type of environment conducive to maintaining star formation. 
 \begin{figure}
 \centering
 \includegraphics[width=7.5 cm,angle=-90]{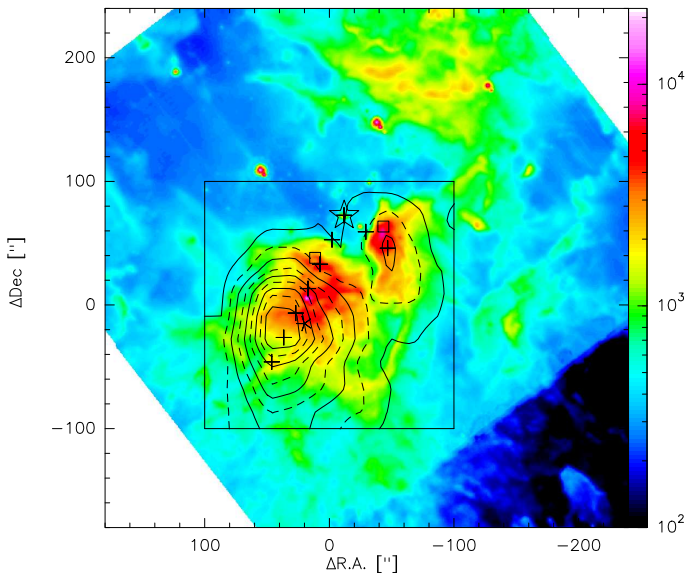}
\caption{ 8~$\mu$m Spitzer/IRAC band (in MJy/sr) overlayed with \mbox{CO 4--3} NANTEN2 observations presented in this paper. The position of the cluster is marked by a star. The boxes mark the positions of the pillar heads. The crosses mark the positions along the cuts into MM1 and MM2.  IRAC pixel size is 1.2". The position (0,0) corresponds to R.A.(J2000.0)=$\rasf{11}{15}{08}{85}$, 
 Dec.(J2000.0)=$\decasf{-61}{16}{50}{0}$.}
\label{spitzer}
\end{figure}
%  \begin{figure*}
%  \begin{minipage}[hbt]{8.5cm}
%  \centering
%  \includegraphics[width=13 cm,angle=-90]{ngc3603_co43_on_spitzer8mu_short.eps}
%  \end{minipage}
%  \hfill
%  \begin{minipage}[hbt]{8.5cm}
%  \centering
%  \includegraphics[width=13 cm,angle=-90]{ngc3603_co43_on_spitzer5.8mu_short.eps}
%  \end{minipage}
% \caption{Spitzer/IRAC and NANTEN2 view of the central parts of NGC~3603. IRAC pixel size is 1.2". {\bf Left:} 8~$\mu$m IRAC band (in MJy/sr) overlayed with CO 4--3 NANTEN2 observations presented in this paper. {\bf Right:} 5.8~$\mu$m IRAC observation  (in MJy/sr) overlayed with NANTEN2 CO 4--3 data. The boxes mark the position of the pillar heads. The crosses mark the positions along the cuts into MM1 and MM2. }\label{spitzer}
% \end{figure*}
On their way into a molecular cloud FUV photons are absorbed, and a depth-dependant chemical balance is established. Models of these so-called photon dominated regions  \citep[PDRs; see references in][]{HT1999, roellig2007} typically predict a stratification of species like H/H$_2$, C$^+$/C/CO, and others. This stratification is found to be independent of the shape of the PDR \citep{gorti02}. However, observations often reveal [CI] emission coincident with that of CO  \citep{tatematsu99, ikeda02, mookerjea2006, sakai06, kramer2008, sun08}. Plane-parallel models cannot explain this behavior unless full face-on orientation is assumed. Spherical models under isotropic illumination show well-correlated [CI] and CO emission for low density clouds ($n\lesssim10^4$cm~$^{-3}$) but predict a limb brightening of the [CI] emission \citep{roellig2006}.
In this work we analyze the distribution of atomic carbon and warm CO in the NGC~3603 star forming region and use LTE approximations and an escape probability model \citep{stutzki85} to derive the excitation conditions of the gas as well as gas abundances. We refrain from applying more detailed models like the KOSMA-$\tau$ PDR model \citep{stoerzer1996, roellig2006, cubick08}. 
Recent Herschel/HIFI observations of massive star forming regions provide additional strict constraints on the PDR modeling of these regions, resulting in significantly altered predictions of the local physical and structural conditions \citep{ossenkopf2010dr21, dedes2010}. Hence, we postpone all attempts of PDR modeling in \mbox{NGC~3603} until the region has been observed with Herschel in the framework of the Herschel Guaranteed Time Key Project {\it Warm And Dense ISM}. 

\section{The NGC 3603 star forming region}
\mbox{NGC 3603} is located in the Carina spiral arm ($l\approx 291.6^\circ, b\approx -0.5^\circ$) at a distance of approximately 7-8 kpc \citep[see discussion in][]{melena2008}. It is one of the most luminous (L$_\mathrm{bol}>10^7$ \lsun), optically visible HII regions in the Galaxy, with the massive OB cluster \mbox{NGC~3603~YC} (Young Cluster) as a power source \citep{goss69}. The compact core of the cluster was designated  HD~97950 due to its star-like appearance. As a comparison, \mbox{NGC~3603} is 100 times more luminous than the Trapezium cluster in Orion.  This young cluster 
can be considered one of the few Galactic starburst clusters, which are essential for the understanding of extragalactic star bursts. 
 With a dynamic stellar
cluster mass of 17600$\pm$3800~M$_\odot$ \citep{rochau2010} residing in a cloud of total gas mass
of $4\times 10^5$ M$_\odot$ \citep{grabelsky88}, it is the most compact 
Galactic star-forming complex outside the Galactic center region \citep{stolte04}.
In a number of recent studies, the stellar cluster and the surrounding HII region has been investigated
thoroughly \citep{pandey2000, sung2004, stolte06, lebouteiller2007,  lebouteiller2008, harayama2008, nuernberger2008, crowther2010}.
To the 
south of the cluster we find a giant molecular cloud (see Figure~\ref{spitzer}). The 
intensive radiation and stellar winds from the cluster shape large gaseous 
pillars at the edge of the cloud \citep{brandner00}. CS observations by \citet{nuernberger02} show that the 
molecular gas also extends much further to the north ($>10$\arcmin) and to the south, hosting numerous massive 
molecular clumps which may be future sites of star formation. 
A number of prominent IR sources have been found so far \citep{frogel77, nuernberger2008, nuernberger2010} as well
as proplyd like objects \citep{muecke2002}. 
Recent AKARI observations of NGC~3603 have been presented by \citet{okada2010} showing that the [CII]~158~$\mu$m 
emission is  widely distributed and that the [OIII]~88~$\mu$m emission follows the MIR, FIR, and radio continuum emission. 
\section{Observations}
 We used the NANTEN2-4m antenna at 4865m altitude in Pampa la Bola in northern Chile 
to map the central  2\arcmin $\times$ 2\arcmin  region of \mbox{NGC~3603}.
Our reference position is R.A.(J2000.0)=$\rasf{11}{15}{08}{85}$, 
Dec.(J2000.0)=$ \decasf{-61}{16}{50}{0}$,
 73\arcsec\  south-east of the
central OB cluster \mbox{NGC~3603~YC} at R.A.(J2000.0)=$\rasf{11}{15}{07}{26}$, 
Dec.(J2000.0)=$ \decasf{-61}{15}{37}{48}$.  
We observed the rotational transitions 
of \mbox{$^{12}$CO  J=4--3} (461.0408~GHz)  and \mbox{J=7--6} (806.6517~GHz) 
and the two fine-structure transitions of atomic carbon \mbox{\CI, $^3$P$_1$--$^3$P$_0$}
(492.1607~GHz) and \mbox{$^3$P$_2$--$^3$P$_1$} (809.3446~GHz) (henceforth 1--0 and 2--1), 
between September and November 2006 with a 
dual-channel 460/810~GHz receiver. The exact line frequencies were taken from 
the {\it Cologne Database for Molecular Spectroscopy} CDMS \citep{mueller05,mueller01}. 
Double sideband (DSB) receiver temperatures
were $\approx$ 250~K in the lower frequency channel and $\approx$ 750~K in
the upper channel. The intermediate frequencies (IF) are
4~GHz and 1.5~GHz, respectively. The latter IF allows simultaneous
observations of the \mbox{CO 7--6} line in the lower
sideband and the \mbox{\CI\ 2--1} line in the upper sideband. These two lines were 
observed simultaneously with one of
the two lower lines in the 460~GHz channel. As backends, we used two
acousto optical spectrometers (AOS) with bandwidths of
1~GHz. The channel spacing was 0.37 km~s$^{-1}$ at 460~GHz
and 0.21 km~s$^{-1}$ at 806~GHz.
The pointing accuracy was checked regularly on Jupiter,
IRC+10216, and IRc2 in OrionA. The applied corrections
were always $<$20\arcsec\ and usually $<$
10\arcsec. To determine the atmospheric transmission, we measured 
the atmospheric emission 
at the reference position. Spectra 
of the two frequency 
bands were calibrated separately,
and sideband imbalances were corrected using the
atmospheric model {\em atm} \citep[Atmospheric Transmission at Microwaves, cf.][]{pardo01}.
Observations
were taken on-the-fly (OTF), scanning in right
ascension at a speed of 2.5\arcsec/sec and sampling every 10\arcsec. The reference position was observed at the
beginning of each OTF scanning line.

The half power beam widths (HPBW) deconvolved
from the observed full widths at half maximum
(FWHM) are 38.0\arcsec and 26.5\arcsec in the lower
and upper receiver bands, respectively. Beam efficiencies $B_\mathrm{eff}$ are  50\% and 45\%, respectively \citep{simon07,kramer2008}.  The median RMS across the map is 1.74, 1.53, 0.70, 
and 1.25~K for \mbox{CO 4--3}, \mbox{7--6}, \mbox{\CI\ 1--0}, and \mbox{2--1} respectively. The forward efficiency, $F_\mathrm{eff}=86$\% in both bands, was determined from sky-dips. The raw data were calibrated to antenna temperatures $T_A^*$ and scaled to main beam temperatures T$_{mb}$ with the factor $F_\mathrm{eff}/B_\mathrm{eff}$. We present all data in units of T$_{mb}$. Fifth order polynomial baselines were subtracted from all spectra.
%MR 
% A few positions required higher order (up to 5th order) baselines to be subtracted.
Observations of the atomic carbon lines are very challenging and require many OTF coverages 
to achieve a signal to noise (S/N) ratio above 3 $\sigma$. 

\section{Data}
\subsection{Dust and PAHs}

In Fig.\ref{spitzer}, we show a map of integrated \mbox{CO
4--3} intensities from NANTEN2 overlayed to an 8~$\mu$m image 
 taken with the Infrared Array
Camera (IRAC) on board the Spitzer space telescope\footnote{Post-BCD 
data were retrieved from the Spitzer archive 
(URL http://archive.spitzer.caltech.edu/). Because of
strong saturation effects, we chose the short exposure data.}. The 
position of the OB cluster is marked by
a star. The image shows prominent molecular clumps southwest and southeast
of the cluster position. Following \cite{nuernberger02}, we refer to
 them as MM1 and MM2, respectively.
The IRAC image shows the sharp PDR interfaces between the HII region and
the molecular cloud. The positions of these interfaces match the 
pillar-like structures
visible in HST and VLT images \citep{brandner00}. Note that these 
pillars are much smaller than the molecular clouds from which they 
protrude. We 
also note a dusty filament, which 
connects MM1 and MM2 in a large arc 1.5' to 2' south of the western 
pillar head. This
filament is not visible in CO or [CI] emission.
% MC: Folgendes sollte man umformulieren:
% The strong 
%  8~$\mu$m emission ($>$ 10$^4$ MJy/sr) is due to the 
%  FUV flux produced by the OB cluster 
% heating the gas and dust in the molecular clouds. The 8\,$\mu$m band, 
% which includes strong PAH emission (7.6,7.8, and 8.6~$\mu$m), traces 
% the PDR interface especially well.
% Vorschlag:
The strong 8~$\mu$m emission ($>$ 10$^4$ MJy/sr) includes strong PAH emission 
(7.6,7.8, and 8.6~$\mu$m).
%, tracing the dust heated by the intense FUV flux.
The 8\,$\mu$m band is an excellent tracer of the PDR interface.
 For a detailed analysis of the emission by PAH's and very
small grains see \citet{lebouteiller2007}.
  
\subsection{NANTEN2 observations}
\subsubsection{Maps of integrated intensity}
\begin{figure}[h]   
  \centering
  \includegraphics[angle=-90,width=8.5cm]{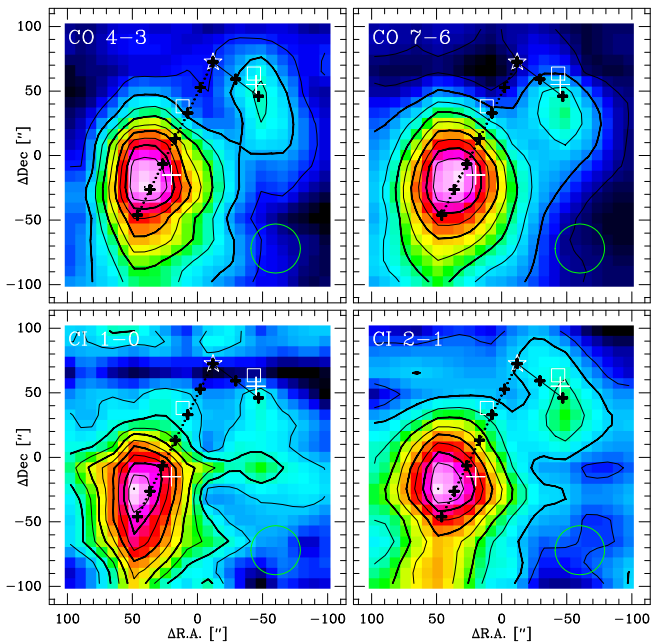}
% Peak integrated intensities:  

\caption{Velocity integrated maps ($200\arcsec\times200\arcsec$) of \mbox{CO
4--3}, \mbox{7--6}, \mbox{\CI\ 1--0}, and \mbox{2--1} smoothed to a common angular resolution of
$38\arcsec$ (1.3-1.5~pc at 7-8~kpc), and integrated over a velocity
range from $4$ to $26$~km~s$^{-1}$. Color scale and contours show the 
same data. Contours range
between 10 and 90\% of the peak intensities which are 
267~K$\cdot$km~s$^{-1}$
for \mbox{CO 4--3}, 121~K$\cdot$km~s$^{-1}$ for \mbox{CO 7--6}, 28~K$\cdot$km~s$^{-1}$ for \mbox{\CI\
1--0}, and 30~K$\cdot$km~s$^{-1}$ for \mbox{\CI\ 2--1}.
The position of the compact OB cluster is marked by a
white star. The position of the the pillars, as seen in the HST images,
are marked by white squares.
Dashed lines and small crosses spaced by $22\arcsec$ mark two cut 
from the OB cluster
to the peak positions of the two clumps MM1 (short cut) and MM2 (long cut).
 The circle marks the resolution and the two white crosses mark 
the position of the 
\mbox{CS(2--1)} peaks \citep{nuernberger02} in MM1 and MM2.  The position (0,0) corresponds to R.A.(J2000.0)=$\rasf{11}{15}{08}{85}$, 
Dec.(J2000.0)=$ \decasf{-61}{16}{50}{0}$.}
\label{fig-ngc3603-4-plots}   
\end{figure}
Fig. \ref{fig-ngc3603-4-plots} shows maps of integrated 
\mbox{CO 4--3}, \mbox{7--6}, \mbox{\CI\ 1--0}, and \mbox{2--1} emission integrated over a velocity
range from $4$ to $26$~km~s$^{-1}$. The two molecular clumps MM1 and MM2 are visible: MM1 to 
the southwest of the OB 
cluster and MM2 to the southeast. 
The peak intensities are 267~K$\cdot$km~s$^{-1}$
for \mbox{CO 4--3}, 121~K$\cdot$km~s$^{-1}$ for \mbox{CO 7--6}, 28~K$\cdot$km~s$^{-1}$ for \mbox{\CI\
1--0}, and 30~K$\cdot$km~s$^{-1}$ for \mbox{\CI\ 2--1}.

 MM2 is very prominent in all 
4 observed transitions
while MM1 is weaker in the fine structure transitions. 
  
\begin{table*}[htb]
\caption{Observed integrated intensities $I=\int T_\mathrm{mb}\,d v$ in units of K$\cdot$km~s$^{-1}$ , v$_\mathrm{LSR}$ in units of km~s$^{-1}$, and the line widths FWHM in units of km~s$^{-1}$ computed as 0$^\mathrm{th}$, 1$^\mathrm{st}$, and 2$^\mathrm{nd}$  moment of the spectra between v$=8...22$~km~s$^{-1}$. The positions lie along a line starting from the OB cluster. The cluster position has been omitted. Rotational transitions of CO are denoted 43 (\mbox{$^{12}$CO 4--3}) and 76 (\mbox{$^{12}$CO 7--6}). The atomic carbon fine-structure transitions are denoted 10 (\mbox{$^3P_1-^3P_0$}) and 21 (\mbox{$^3P_2-^3P_1$}). 
%The position 20\arcsec/-20\arcsec corresponds to the position of MM2, derived from 
%CS (2--1)\citep{nuernberger02}. 
}\label{tabobs} 
\centering  
\begin{tabular}{c|c|c|c|c|c|c|c|c|c|c|c|c}
\hline \hline
 $\Delta\alpha/\Delta\delta$ & $I_{43}$ & $\Delta$v$_{43}$ & v$_\mathrm{LSR}$& $I_{76}$ & $\Delta $v$_{76}$ & v$_\mathrm{LSR}$& $I_{10}$ & $\Delta$v$_{10}$& v$_\mathrm{LSR}$&$I_{21}$&$\Delta$v$_{21}$& v$_\mathrm{LSR}$ \\

 [ \arcsec /\arcsec ]&  [K$\cdot$km/s] & [km/s]& [km/s]& [K$\cdot$km/s] &  [km/s]&  [km/s]& [K$\cdot$km/s] & [km/s]& [km/s]& [K$\cdot$km/s] & [km/s]& [km/s]\\  \hline
{10/30} & 48 & 3.5 & 13.2 & 29 & 8.2 & 13.9 & 6 & 7.4 & 14.5 & 7 & 10.0 & 15.7 \\
 {20/10} & 153 & 6.7 & 14.2 & 71 & 7.4 & 13.5 & 12 & 8.2 & 13.6 & 14 & 7.8 & 15.0 \\
 {30/-10} & 252 & 7.8 & 14.9 & 119 & 7.5 & 13.6 & 27 & 7.1 & 14.6 & 24 & 6.1 & 14.6 \\
 {40/-30} & 267 & 8.4 & 15.2 & 121 & 7.8 & 13.7 & 27 & 5.2 & 14.2 & 30 & 7.1 & 14.1 \\
 {50/-50} & 175 & 7.1 & 14.8 & 84 & 7.7 & 13.8 & 26 & 7.2 & 14.3 & 20 & 6.4 & 14.3 \\
 %{20/-20} & 205 & 7.1 & 14.5 & 106 & 7.5 & 13.3 & 13 & 1.6 & 13.2 & 22 & 6.6 & 14.3 \\
% {-10/70} & 20 & 8.0 & 12.2 & 7 & 9.7 & 13.7 & -4 & 16.2 & 13.8 & 5 & 8.1 & 17.4 \\
\hline {-30/60} & 57 & 4.9 & 15.0 & 24 & 7.5 & 13.0 & 0 & 0.0 & 50.1 & 6 & 3.7 & 16.3 \\
 {-50/50} & 90 & 6.0 & 13.8 & 40 & 7.7 & 13.3 & 11 & 8.4 & 13.5 & 9 & 6.9 & 14.7\\

 \hline \hline
\end{tabular}
\end{table*}

Both CO maps mark the transition from the HII region to the molecular 
cloud. MM1 is presumably smaller than our beam size 
\citep{nuernberger02}, hence it is not possible to infer detailed structure 
information from the maps.

%The atomic carbon maps do not show an equally
%steep transition from the HII region to the molecular cloud and appear
%to be more widespread  compared to CO. 
% MC: Formulierung würde ich etwas ändern.
% The atomic carbon maps do not show an equally
% visible transition regions. It is known that \CI\ observations often appear 
% to be much more widespread compared to CO, however, the S/N of both \CI\ transitions 
% is too bad to allow similar conclusions.
% Vorschlag:
The observed intensity distributions of the \CI\ fine 
structure transitions
are more extended with a lower contrast in comparison 
to the intensity distributions
of the CO rotational line transitions.
The peak positions of [CI] are slightly shifted away from the cluster compared to CO.
However, the S/N of both \CI\ transitions 
is too low to allow further conclusions.
 MM1 can be identified in both CO maps. In \CI\ a diffuse emission
is visible without a clear distinction between clump center and its more diffuse 
environment. 
 The MM2 peak positions of both CO maps match within the pointing accuracy. The same is true for 
 both \CI\ maps. Both \CI\ peaks are shifted away from the OB cluster 
with respect to CO.
% This shift    and \CI\ 2--1 match very well, while the
%peak of the \CI\ 1--0 emission is shifted (~15\arcsec) to the south-east.  
However,  we note a  good spatial correlation between the CO and
 \CI\ emission as all peaks coincide within a beam radius. We
do not see any C-CO layering as expected from simple 
edge-on, plane-parallel PDR scenarios. 
It is possible to explain coincident \CI\ and CO emission
 with a face-on configuration,
but the visible pillar structures in the HST observations 
and the very sharp interface
visible in the IRAC image in Fig.\ref{spitzer} clearly 
show a prominent, edge-on 
 PDR interface. 

\begin{figure}[h]   
  \centering
  \includegraphics[angle=-90,width=8.5cm]{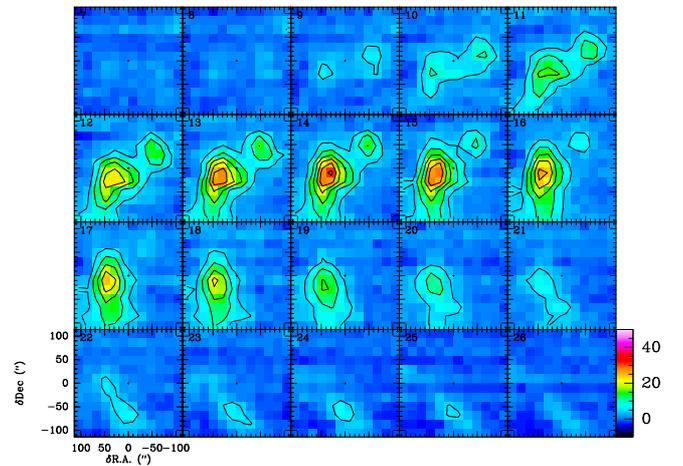}
\caption{Velocity structure of the observed field: \mbox{CO 4--3} intensity in velocity
channels of 1~km~s$^{-1}$ width.  Contours range between 5 and
45~K in steps of 5~K.  The position (0,0) corresponds to R.A.(J2000.0)=$\rasf{11}{15}{08}{85}$, 
Dec.(J2000.0)=$ \decasf{-61}{16}{50}{0}$. }
\label{fig-ngc3603-co43-channels}   
\end{figure}

The velocity structure of the observed field is shown in the \mbox{CO 4--3} velocity
channel maps (Fig.\,\ref{fig-ngc3603-co43-channels}). From the channel
 maps, it is evident
that the emission towards MM1 shows a smaller kinematic range than in MM2. 
MM1 is not 
observable at velocities higher than 16~km~s$^{-1}$, while 
the emission in MM2 extends up
to 21~km~s$^{-1}$. At larger velocities, the emission gradually shifts to the
southwest.
 A similar kinematic behavior is also visible from CS (3-2) observations 
by \citet{nuernberger02}. Their results also show a velocity drift for
MM1 with distance from the OB cluster, possibly indicating a placement of
MM1 in front of the cluster. 
In the appendix we show the velocity channel maps of the
remaining 3 transitions.
\subsubsection{Spectra along a cut through the region}
\begin{figure}[h]   
  \centering   
  \includegraphics[angle=-90,width=8.5cm]{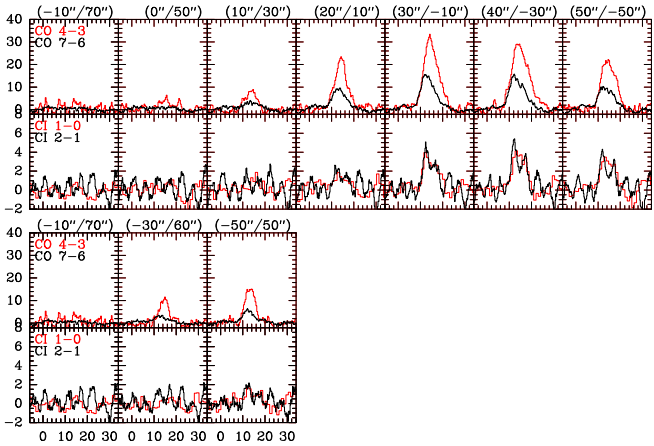}
  \caption{Spectra along two cuts connecting
    the OB cluster and the centre of MM1 (bottom) and MM2 (top).
 The CI spectra appear to show sinusoidal standing waves, 
however a 
detailed analysis shows no consistent standing wave pattern. These  
baseline variations are thus attributed to instabilities of the system and/or the 
atmosphere during the measurements. 
    All data are at a
    common resolution of $38\arcsec$ and on the main beam 
temperature scale.} 
\label{fig-ngc3603-cut-spectra}   
\end{figure} 
To study the transition from the ionized to the molecular gas, we selected 
two cuts from the OB cluster to the peak positions of MM1 and MM2. The positions
along the cut are 22\arcsec\ apart and marked in Fig.~\ref{fig-ngc3603-4-plots}.  
The spectra along the two cuts, 
shown in Fig.\ref{fig-ngc3603-cut-spectra},
 trace the kinematic
structure of the interface regions and the molecular clumps. 
The spectra at the positions \mbox{30\arcsec/-10\arcsec}, \mbox{40\arcsec/-30\arcsec}, and \mbox{50\arcsec/-50\arcsec} 
do not show simple Gaussian 
line profiles. \cite{nuernberger02} observed similar line profiles in their
\mbox{CS 3--2} and \mbox{2--1} observations of MM2. They used two Gaussian components
separated by 3~km~s$^{-1}$ to fit the emission profiles. Our analysis
showed that it is not possible to use a similar two component 
Gaussian fit to reproduce our observations at all positions. 
We derived the first three moments of the spectral line, i.e., 
integrated intensity, mean velocity, and FWHM, over the channels in the range 
$v=6-22$~km~s$^{-1}$.
The derived moments are
given in  Table \ref{tabobs}.  At the offsets -10\arcsec/70\arcsec and 0\arcsec/50\arcsec we don't find
sufficient emission to derive significant values.

\begin{table*}[htb]
\caption{Line ratios R$_{74}$ ($I_{76}/I_{43}$), R$_{21}$ ($I_{21}/I_{10}$) and R$_{14}$ ($I_{10}/I_{43}$) for all positions along the two cuts. The LTE excitation temperature T$_\mathrm{ex,43}$ is derived from the optically thick $I_{43}$ emission, T$_\mathrm{ex,74}$ is derived from R$_{74}$  The C column densities are derived assuming LTE and optically thin \CI\ emission. Values of the integrated intensities of \mbox{C$^{18}$O 2--1} are given as presented in \cite{nuernberger02} along the two cuts, the LTE excitation temperature is derived from $I_{43}$. The LTE column densities of C$^{18}$O, and the total gas column density assume an isotope ratio of 500 and CO:H$_2$=8$\times10^{-5}$ \citep{frerking82}, i.e. a ratio of C$^{18}$O/H$_2$=2$\times10^{-7}$.}\label{tablte}
\centering
\begin{tabular}{c|c|c|c|c|c|c|c|c|c|c|c|c}
 \hline \hline
 $\Delta\alpha/\Delta\delta$ &  R$_{74}$ &R$_{21}$ & R$_{14}$&T$_\mathrm{ex,43}$&T$_\mathrm{ex,74}$&  N$_\mathrm{C}$&$\int\,T_\mathrm{mb}(\mathrm{C}^{18}\mathrm{O})dv$ & N$_\mathrm{C^{18}O}$&N$_\mathrm{tot}$&N(C)/N$_\mathrm{tot}$&M$_\mathrm{LTE}$&M$_\mathrm{vir}$\\

 [\arcsec / \arcsec ]  & & & &[K]&[K]&cm$^{-2}$ & [K km s$^{-1}]$ &[10$^{15}$cm$^{-2}]$&[10$^{21}$cm$^{-2}]$&&[M$_\odot$]&[M$_\odot$] \\ \hline
{10/30} & 0.62 & 1.15 & 0.14 & 23 &62& 9.5$\times 10^{16}$ & 1.6 & 2.27 & 11.35 & 8.4$\times 10^{{-6}}$ & 236 & 208 \\
 {20/10} & 0.47 & 1.09 & 0.08 & 33 &53& 1.7$\times 10^{17}$ & 1.9 & 3.57 & 17.83 & 9.7$\times 10^{{-6}}$ & 371 & 751 \\
 {30/-10} & 0.47 & 0.91 & 0.11 & 43 &53& 3.9$\times 10^{17}$ & 4.2 & 9.84 & 49.19 & 7.9$\times 10^{{-6}}$ & 1024 & 1006 \\
 {40/-30} & 0.45 & 1.09 & 0.10 & 42 &52& 4.1$\times 10^{17}$ & 4.3 & 9.87 & 49.36 & 8.2$\times 10^{{-6}}$ & 1028 & 1175 \\
 {50/-50} & 0.48 & 0.79 & 0.15 & 34 &54& 3.7$\times 10^{17}$ & 2.1 & 4.04 & 20.20 & 1.9$\times 10^{{-5}}$ & 421 & 845 \\\hline
% {20/-20} & 0.52 & 1.62 & 0.07 & 39 &56& 2.2$\times 10^{17}$ & 4.7 & 10.13 & 50.66 & 4.3$\times 10^{{-6}}$ & 1055 & 845 \\
% {-10/70} & 0.39 & -1.31 & -0.20 & 10 &48& -1.1$\times 10^{17}$ & 0.7 & 0.61 & 3.06 & -3.6$\times 10^{{-5}}$ & 64 & 1072 \\
 {-30/60} & 0.43 & 8.89 & 0.01 & 21 &51& 1.$\times 10^{17}$ & 1.3 & 1.73 & 8.64 & 1.2$\times 10^{{-5}}$ & 180 & 399 \\
 {-50/50} & 0.44 & 0.84 & 0.12 & 24 &51& 1.5$\times 10^{17}$ & 2.1 & 3.07 & 15.37 & 1.$\times 10^{{-5}}$ & 320 & 599 \\
\hline \hline
\end{tabular}
\end{table*}
% \begin{table}[htb]
% \caption{Values of the integrated intensities of C$^{18}$O 2--1 as presented in \cite{nuernberger02} along the two cuts, the LTE excitation temperature as derived from $I_{43}$, the LTE column densities of C$^{18}$O, and the total gas column density, assuming an isotope ratio of 500 and CO:H$_2$=8$\times10^{-5}$ \citep{frerking82}, i.e. an assumed ratio of C$^{18}$O/H$_2$=2$\times10^{-7}$.
% }\label{columnlte}
% \centering
% \begin{tabular}{c|c|c|c|c|c}
%  \hline \hline
% $\Delta \alpha $/$\Delta \delta $&$\int\,T_\mathrm{mb}dv$ & T$_\mathrm{ex}$& N$_\mathrm{C18O}$&N$_\mathrm{tot}$&N(C)/N$_\mathrm{tot}$ \\

%  [ \arcsec /\arcsec ] & [K km s$^{-1}]$ &[K]&[10$^{15}$cm$^{-2}]$&[10$^{21}$cm$^{-2}]$&\\\hline
% %-10/70 & 0.7 & -- &  -- & --& -- \\ 
% %0/50 & 1.5 & 10 & 1.31 & 6.55& -- \\
% 10/30 & 1.6 & 18 & 1.89 & 9.47& 1.4$\times 10^{-5}$  \\
% 20/10 & 1.9 & 32 & 3.5 & 17.5&1.1$\times 10^{-5}$  \\
% 30/-10 & 4.2 & 46 & 10.4 & 51.8&6.9$\times 10^{-6}$  \\
% 40/-30 & 4.3 & 47 & 10.8 & 54.1&7.8$\times 10^{-6}$  \\
% 50/-50 & 2.1 & 41 & 4.75 & 23.7&1.5$\times 10^{-5}$  \\
% 20/-20 & 4.7 & 46 & 11.7 & 58.6&3.2$\times 10^{-6}$  \\
% %-10/70 & 0.7 & --&--&--&\\
% -30/60 & 1.3 & 15 & 1.41 & 7.05&1.4$\times 10^{-5}$  \\
% -50/50 & 2.1 & 26 & 3.23 & 16.2&9.3$\times 10^{-6}$  \\
% \hline \hline
% \end{tabular}
% \end{table}
\subsection{Analysis}
\subsubsection{LTE -- Column densities and temperatures}
In Table \ref{tablte} we give the velocity integrated line ratios R$_{74}$ ($I_{76}/I_{43}$),
 R$_{21}$ ($I_{21}/I_{10}$) and R$_{14}$ ($I_{10}/I_{43}$) for all 
positions along the 
two cuts. We compute the LTE temperatures using the optically thick
\mbox{CO 4--3} emission assuming a beam filling of 1 and obtain temperatures of 42~K for the 
peak position in
MM2 and 24~K for MM1. This is roughly consistent with estimates 
from the \CI\ line ratio.  R$_{21}$ is a 
sensitive function of the \CI\
excitation temperature.  In the optically thin limit and assuming LTE,
$T_{\rm ex}=38.3\,{\rm K}/\ln[2.11/R_{21}]$. In MM2 we 
find ratios between 
0.8 and 1.1
which correspond to temperatures between 40 and 60~K.
For MM1 the \CI\ line ratio 
gives a temperature estimate of 42~K, higher than the estimate derived 
from the CO 4-3 emission. The difference can be explained by beam dilution
 effects since MM1 is not resolved within the NANTEN2 beam. However, the
 atomic carbon emission is heavily affected by noise. Another possibility 
is to use the LTE ratio of the two optically thick lines $R_{74}$ to derive an 
excitation temperature: $T_{\rm ex}=99\,{\rm K}/\ln[3.06/R_{74}]$. $R_{74}$ 
is remarkably constant across both clouds. We find an excitation temperature
 slightly above 50~K for both clouds. First, this confirms that both clouds 
have comparable excitation conditions despite their different appearance.
 Second, they seem not to be composed of hot, unresolved clumps which would result in a much higher excitation temperature derived from  $R_{74}$. Instead 
the gas appears to be more smoothly distributed and excited.

The atomic carbon column density was derived under the assumption of LTE
and optically thin \CI\ emission. 
 To derive the total H$_2$ column densities we use the integrated
 intensity of \mbox{C$^{18}$O 2--1} as presented in \cite{nuernberger02}.
 Assuming optically thin emission, an isotope ratio of 
500 and CO:H$_2$=8$\times10^{-5}$ 
\citep{frerking82}, i.e., an assumed ratio of
 C$^{18}$O/H$_2$=2$\times10^{-7}$, we compute the column densities
 along the two cuts.
The results are summarized in Table \ref{tablte}. 
For MM1, we find a total column density of 1.5$\times10^{22}$~cm$^{-2}$;
 for MM2, $N=2...5\times10^{22}$~cm$^{-2}$.  We find that the 
relative abundance of atomic carbon  N$_\mathrm{C}$/N$_\mathrm{tot}$ drops towards the peak positions
with values of  $10^{-5}$ (MM1) and $8-19\times10^{-6}$ (MM2), and 
increases at positions towards as well as away from the OB cluster.
Even though the spatial resolution of our maps is rather coarse,
this could be an indication for a C/CO stratification, because in this case
we would expect limb brightened [CI] emission.
The gas pressure within  MM1 and MM2 (assuming a density of 10$^4$~cm$^{-3}$) is 
about 10-30\% of the pressure in the HII region \citep{shaver70}. Hence, the
interface region is compressed and driven against the molecular 
cloud, visible, e.g., in the pillars protruding from MM1 and MM2.  
From the total column density we can estimate the total mass per beam. The 
results are given in Table~\ref{tablte}. We also derived virial mass estimates
across the cuts assuming a density of
10$^4$~cm$^{-3}$ and kinetic temperatures of 50~K (the densities are in agreement
with the calculations from section~\ref{chapterEP}). The virial masses agree 
very well with the LTE masses and are much 
larger than the Jeans mass of the material of 87 M$\odot$. This means, the gas is 
gravitationally bound and not transient. We expect that the strong
HII pressure and the gravitational instability causes  strong fragmentation and 
cloud collapse.  \cite{nuernberger02} derived a star formation efficiency of $\ge 30\%$
 and a mean star formation rate of $1.3\times 10^{-3}$~M$_\odot$/yr.  Embedded IR 
sources have already been identified \citep{frogel77} along the outskirts of the 
molecular clouds and particularly at the base of the pillar like structures at
the edge of MM2 \citep{nuernberger2008, nuernberger2010}.

 \begin{figure*}
 \begin{minipage}[hbt]{8.5cm}
 \centering
  \includegraphics[angle=-90,width=8.5cm]{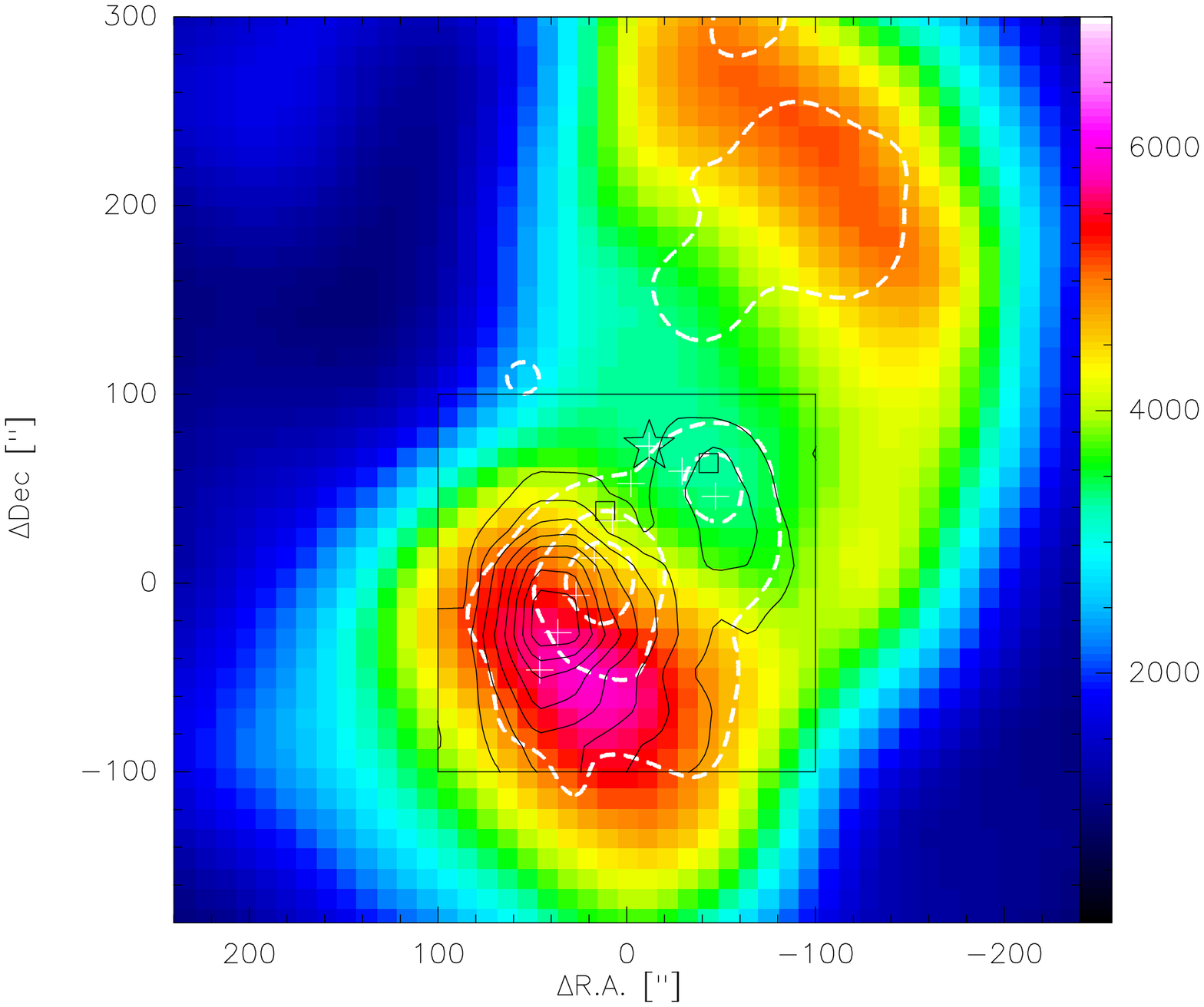}

 \end{minipage}
 \hfill
 \begin{minipage}[hbt]{8.5cm}
 \centering
 \includegraphics[width=7 cm,angle=-90]{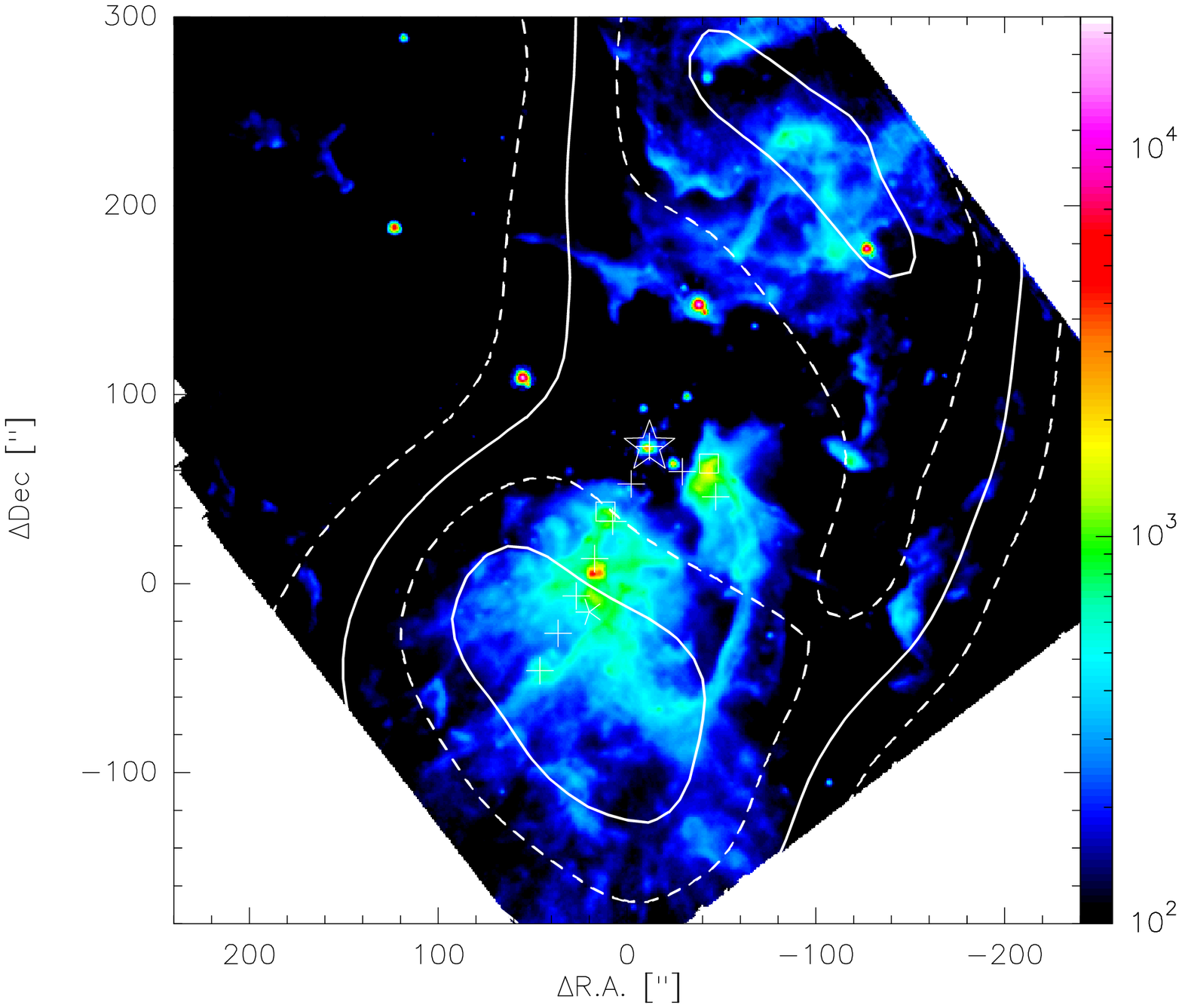}
 \end{minipage}
\caption{{\bf Left:}Values of the FUV field  in units of the Draine field derived from the HIRES 60, and 100~$\mu$m
 fluxes as color map. White, dashed contours mark the Spitzer/IRAC 5.8$\mu$m flux smoothed to a resolution 
of 35". The contour values are 25, 50, and 75\% of the peak value. Overlayed in thin contours 
 is the NANTEN2 \mbox{CO 4--3} map. The central OB cluster is marked by a star.  The two pillar heads 
are denoted by the squares, White crosses mark the positions along the cuts into MM1 and MM2.
{\bf Right:}Spitzer/IRAC 5.8~$\mu$m observation  (in MJy/sr) overlayed with FUV fluxes as derived from IRAS/HIRES flux ratios. The contours correspond to values of $\chi$\,=\,2000, 3000, 4000, and 5000. IRAC pixel size is 1.2".  The position (0,0) in both panels corresponds to R.A.(J2000.0)=$\rasf{11}{15}{08}{85}$, 
Dec.(J2000.0)=$ \decasf{-61}{16}{50}{0}$.}\label{fig-ngc3603-fuv}
\end{figure*}

\subsubsection{Dust Temperatures and FUV intensities}
%\begin{figure*}[hbt]   
%  \centering   
%  \includegraphics[angle=0,width=15cm]{ngc3603_escprob_paper.eps}
%  \caption{Comparison of the observed line ratios $^{12}$CO7--6/$^{12}$CO4--3, \CI 1--0 /$^{12}$CO4--3, and \CI 2--1/ \CI 1--0 with results from an escape probability model at the peak positions of MM1 and MM2. The middle, solid contours represent the observed ratios, the outer dash-dotted contours represent the 20\% uncertainties. The grey-scale image shows the reduced $\chi^2$ of the fit. The white, dotted contours correspond to $\chi^2$ values of 1, 2, and 5.}  
%\label{fig-ngc3603-escprob}   
%\end{figure*}   
To estimate  the total FUV flux, we consider the 
total luminosities of the
most massive O stars in the central OB cluster.
 Recently, \citet{melena2008} published an updated census of
the massive star content of NGC~3603. 
The most
massive members of the OB cluster are 3 WNL stars, 
14 O3( and 3.5) stars (III+V) and at least 20 late O-type stars. 
They produce 
the FUV flux that dominates the heating of the PDRs and the 
molecular clouds.  At an 
effective temperature of 24000~K, the ratio of the
 FUV energy density to the total energy density emitted by the star
 $\Phi_\mathrm{FUV}/\Phi_\mathrm{tot}\approx0.7$ is maximal 
if we assume pure black body emission. At higher temperature, 
relatively more energy is emitted in the EUV range
 ($13.6\,\mathrm{eV}< h\nu < 130\,\mathrm{eV}$)\footnote{
\citet{pauldrach98} showed in model calculations that for 
effective temperatures below 45000~K, the ratio of FUV to EUV 
photon rates is considerably higher than the ratio derived for 
pure blackbody spectra \citep[see also][]{brandner00}. This
is because EUV photons are absorbed in the ionization front.
We neglect this effect.}.

Applying effective temperatures and luminosities as given by
\citet{martins2005} for the O stars, 
 \citet{panagia73} for the B stars  and \citet{crowther2007} for the WR stars, we calculated a total FUV luminosity of
 $1.1\times10^{43}$~erg~s$^{-1}$ from the contribution of each cluster member. This is a lower limit since the spectral classes for more than 10 cluster members remains unknown.  Depending on the distance from the cluster center $d$, in units of
 pc, the FUV flux is $\chi=9.4\times10^4 d^{-2}$ in units of the Draine field. At a distance of 7-8~kpc,
 one parsec corresponds to angular distances of 26-29\arcsec, i.e., somewhat smaller
 than the projected distance between the cluster center and the edge of MM2 as seen
 for example in Fig.~\ref{spitzer} ($\approx 46$\arcsec). If we assume
 that the cluster and the molecular cloud are situated in the same
 plane, the FUV field at the peak position in MM2 drops to $\chi\approx 5000-6600$.
 This is of the same order as the flux at the MM2-peak derived from the 
IRAS data. If the clumps are displaced from the plane of the OB cluster, or 
if the distance to the complex is different, the derived $\chi$ changes.
Uncertainties 
in the in-plane displacement of $\pm$ 4 pc and in the distance (7-8 kpc) lead to  $\chi=2400-6600$.

We use IRAS data to derive dust temperatures and the FUV continuum in the observed region.
 Enhanced
  resolution images of $\sim1'$ resolution were created using the
  maximum correlation method \citep{aumann90}.
We estimate dust temperatures from the ratio of HIRES
60~$\mu$m and 100~$\mu$m data, assuming a dust spectral index of emissivity
 of 1.5. We obtained 
 high resolution (HIRES) 60
  and 100~$\mu$m images (2\degree$\times$2\degree) from the IPAC data center\footnote{http://www.ipac.caltech.edu/}.
  Along the two cuts, we find almost 
constant dust temperatures of 35~K.  This value is very close to the dust temperature
of 37~K derived from MSX data by \citet{wang2010}. 
Again, comparison with the higher resolution Spitzer maps shows that the warm
 dust is not distributed homogeneously. We conclude that the HIRES data are, 
therefore, affected by beam filling.
   
 Following \cite{nakagawa98},
  we combined 60 and 100$\mu$m data to create a map of
  far-infrared intensities between 42.5~$\mu$m
  and 122.5~$\mu$m \citep{helou88}. Under the assumption that all
  FUV energy absorbed by the grains is re-radiated in the far-infrared (FIR),
  we compute the FUV fluxes
  ($\chi$) from the emergent FIR
  intensities \citep[cf.][]{kramer05} $I_{\rm FIR}$, using $\chi/\chi_0=4\pi~I_{\rm FIR}$, with
 $\chi_0=2.7\times 10^{-3}$~erg~s$^{-1}$~cm$^{-2}$ 
\citep{draine78}.  The resulting map of the FUV continuum is shown in the left panel of
Fig. \ref{fig-ngc3603-fuv}.  Overlayed as thin contours 
is the NANTEN2 CO 4--3 map. At 
the peak position of
MM1, we find FUV fields of $\chi=3460$. Along the cut in MM2, we 
find that $\chi$ gradually increases from 3350 at the position of 
the OB cluster 
to 5230 at the peak position. Due to the low spatial resolution
of the HIRES data, these values are lower limits since beam dilution effects
are not negligible. 
   
\begin{table*}[htb]
\caption{Results from the escape probability calculations. The  filling factor  $f=I_\mathrm{obs}$/ I$_\mathrm{mod}$ is derived from the CO 4-3 intensities. The CO and C column densities are results from the escape probability fit. }\label{tabep}
\centering
\begin{tabular}{rrrrrrrrrr}
 \hline \hline
{$\Delta \alpha $/$\Delta \delta $ } & T & n & $f$ &{\centering N$_\mathrm{CO}$ } & N$_\mathrm{C}$ & N$_\mathrm{tot}$ &{\centering N$_\mathrm{tot}\times f$} & M$_\mathrm{EP}$ & $<$n$>$ \\
 {[$\texttt{"}$]/[$\texttt{"}$]} & [K] & [{cm}$^{-3}$] &  & [10$^{17}${cm}$^{-2}$] & [10$^{17}${cm}$^{-2}$]
& [10$^{21}${cm}$^{-2}$]& [10$^{21}${cm}$^{-2}$] & [M$_{\odot }$] & [{cm}$^{-3}${/beam]} \\
\hline
% {-10./70.} & 25 & 1.$\times$ 10$^3$ & 0.050 & 139.00 & 0.0 & 163.0 &8.1& 3399 & 47948 \\
% {0./50.} & 39 & 6.8$\times$ 10$^3$ & 0.108 & 56.10 & 5.8 & 66.0 &7.1& 1374 & 19382 \\
 {10./30.} & 53 & 3.$\times$ 10$^3$ & 0.112 & 28.3 & 4.6 & 33.3 &3.7& 693 & 9775 \\
 {20./10.} & 48 & 5.2$\times$ 10$^3$ & 0.277 & 23.7 & 5.8 & 27.9 &7.7& 581 & 8197 \\
 {30./-10.} & 49 & 4.5$\times$ 10$^3$ & 0.429 & 43.3 & 12.3 & 51.0 &21.8& 1061 & 14969 \\
 {40./-30.} & 47 & 1.6$\times$ 10$^4$ & 0.665 & 32.3 & 12.4 & 38.0 &25.3& 791 & 11154 \\
 {50./-50.} & 42 & 9.2$\times$ 10$^3$ & 0.383 & 19.5 & 12.3 & 22.9 &8.8& 477 & 6734 \\ \hline
% {20./-20.} & 57 & 1.$\times$ 10$^7$ & 1.090 & 19.4 & 11.5 & 22.8 &24.8& 474 & 6688 \\
% {-10./70.} & 25 & 1.$\times$ 10$^3$ & 0.050 & 139.0 & 0.0 & 163.0 &8.1& 3399 & 47948 \\
 {-30./60.} & 65 & 3.$\times$ 10$^3$ & 0.125 & 26.8 & 9.0 & 31.5 &4.0& 657 & 9262 \\
 {-50./50.} & 43 & 3.$\times$ 10$^3$ & 0.187 & 38.5 & 6.1 & 45.2 &8.5& 942 & 13288\\
\hline \hline
\end{tabular}
\end{table*}

One would expect a decrease of FUV with growing distance from the cluster. 
However, such a decrease is not
 seen, Fig.~\ref{fig-ngc3603-fuv} shows the opposite.
%MR Volkers Kommentare eingebaut
This is because the method of deriving FUV intensities from IRAS data is
not working for too small dust column densities because then UV radiation can not 
be effectively transformed to IR. Consequently, it would be best to 
exclude these parts from the further analysis. In our case, this is 
impossible due to the large beam. HIRES beam sizes vary from
 $\sim$35\arcsec to over 120\arcsec across the maps, hence the FUV
 values in Fig.~\ref{fig-ngc3603-fuv} are considerably beam
 diluted. In the right panel in Fig.~\ref{fig-ngc3603-fuv}, we overlayed
 the FUV fields derived from the HIRES data on 5.8$\mu m$ Spitzer/IRAC observations
of \mbox{NGC~3603} to illustrate the beam filling effect:
% The FUV map was derived from HIRES data measured at 
60 and 100$\mu$m emission can only be measured if dust is present at all.
 The Spitzer map shows no dust emission between the interface and the
 cluster. The HIRES beam size at the cluster position is almost circular
 with a 
FWHM of $\sim 35$\arcsec.  Hence, a contribution to the IRAS/HIRES
 fluxes at the cluster position can only come from the interface regions of MM1
 and MM2.
If 60/100$\mu$m emission could be measured at 1-2\arcsec resolution, one 
could not determine the FUV field towards the cluster and would observe a sharp
interface between the \HII\ region and the clumps. Instead we see no zero
FUV intensity at the cluster and a gradual increase up to the peak 80\arcsec\ 
behind the interface.
 This is further illustrated
in the left panel in Fig.~\ref{fig-ngc3603-fuv} where we additionally overlayed
Spitzer/IRAC 5.8$\mu$m emission smoothed to the best HIRES resolution of 
35\arcsec. At this resolution, up to 50\% of the total flux is smeared 
across the \HII\  region.
%
% The derived FUV
% fluxes at the cluster position are a factor 20 smaller than the estimates
% from the spectral types.

 At the MM2-peak, the HIRES beam is almost fully filled
 and the FUV fluxes are consistent with the assumption that the molecular
 cloud and the cluster lies in almost the same plane of the sky. This also supports the 
assumption that the FUV irradiation of the OB cluster created the pillars seen in the HST/VLT images. 

%The above analysis is also consistent with the FUV field determined from 
%[CII] fluxes measured by ISO satellite.
The derived FUV fluxes can be used to estimate the ratio $I_\mathrm{CII}/I_\mathrm{FUV}$,
i.e. the fraction of FUV energy that is re-emitted in the the \mbox{[CII] 158~$\mu$m} 
cooling line. 
 We retrieved the [CII] spectrum\footnote{Taken at R.A.(J2000.0)= $\rasf{11}{15}{10}{2}$, Dec.(J2000.0)= $\decasf{-61}{16}{45}{52}$, FITS file name: lsan20100302.fits}  
from the ISO data archive\footnote{http://isowww.estec.esa.nl/ida/}
 and compute an integrated intensity of 
$2\times10^{-3}$erg~s$^{-1}$~cm$^{-2}$~sr$^{-1}$.
Most of the FUV energy that irradiates the cloud is re-radiated in the 
continuum, but a small fraction
is re-emitted in cooling lines, such as the \mbox{[CII] 158$\mu$m} line. 
Applying the FUV flux derived from the HIRES data, we estimate 
$I_\mathrm{CII}/I_\mathrm{FUV}\approx 0.2..0.3\%$. This is consistent with
similar results from regions of active star formation, e.g., nuclear regions of galaxies, 
where approximately 0.1 to 1 \% of the total 
FUV input is re-radiated in the [CII] line \citep{stacey91}. However, this remains a rough 
estimate because of the large beam sizes involved.
% and the unknown contribution of the 
%HII region to the measured CII flux. 

Considering the similar FUV estimates from stellar luminosities and from the FIR data, we 
conclude that the FUV flux illuminating the molecular 
clumps MM1 and MM2 is 3000-6000 in units of the Draine field. Approximately 0.2 to 0.3\% of the total FUV energy 
is re-emitted in the [CII] 158$\mu$m cooling line.

\subsubsection{Escape Probability -- density and temperatures}\label{chapterEP}
We performed escape probability (EP) calculations \citep{stutzki85} to derive local gas densities, column densities, and gas temperatures from the observed line ratios (R$_{74}$,R$_{21}$,R$_{14}$, and $I_\mathrm{C^{18}O 2-1}/I_{43}$). The model assumes a finite, spherical cloud geometry. We fitted the column densities of $^{12}$CO and C independently. From the absolute \mbox{CO 4--3} intensities we derived a filling factor I$_\mathrm{obs}$/ I$_\mathrm{mod}$. The total H$_2$ column densities and masses have been calculated using CO:H$_2$=\mbox{$8\times 10^{-5}$}.  The results are shown in Table~\ref{tabep}.

For MM2, the temperatures from the EP calculations agree within 25\% with the LTE temperatures for the positions \mbox{30\arcsec/-10\arcsec},  \mbox{40\arcsec/-30\arcsec}, and  \mbox{50\arcsec/-50\arcsec}.
%or the two positions closest to the interface, the EP temperatures are higher by 50-130\%. 
% The LTE column densities are higher because no filling factor was applied. Apart from that, both column densities are comparable. That applies to the masses too.
Masses and column densities are comparable with results from the LTE calculations.
 The filling factor is roughly 0.3-0.6 across MM2. MM1 shows significantly smaller values of $~0.1-0.2$. This is consistent with the morphology namely that the clouds are clumpy and MM1 is not resolved by the NANTEN2 beam.
%Consequently, the column densities and masses are also significantly smaller compared to the values in Table~\ref{tablte}. 
The derived temperatures for MM1 are about a factor 2 larger compared to the temperatures derived from \mbox{CO 4--3} and are compatible with the temperatures derived from the observed ratio R$_{74}$ (consistent with $f<1$). The gas densities found for MM1 and MM2 are about $10^4$ cm$^{-3}$. In contrast to the LTE results, we find a monotonously growing N$_\mathrm{C}$/N$_\mathrm{H_2}$ along the cut in MM2 with values of 1-5 $\times10^{-5}$.  
 The column densities, given in Table~\ref{tabep}, result in  C/CO abundance ratios between 0.16--0.6 Similar abundance ratios have been observed in other Galactic star forming regions like CepheusB, Orion or NGC7023 \citep[see references in][]{mookerjea2006}. 
 The C/CO ratio is supposed to depend on the local FUV intensity due to dependency of the CO formation and destruction balance on the FUV illumination. \citet{sun08} found significantly lower C/CO ratios in IC348 where the FUV field varies between 1 and 100 Draine units. On the other hand, \citet{kramer2008} found  C/CO ratios comparable to our results in the Carina region, where the FUV field is also comparable to \mbox{NGC~3603}.
We also find a monotonously growing C/CO abundance ratio along the cut in MM2. A similar trend has been found by \citet{mookerjea2006}, namely that the C/CO ratio is higher further away from the sources of FUV radiation.

\section{Summary and Conclusions}
We used the NANTEN2-4m telescope to map the emission of atomic carbon and CO in the vicinity of the central OB cluster in the Galactic star forming region \mbox{NGC~3603}. These data are the first observations of \mbox{CO 4--3}, \mbox{CO 7--6}, \mbox{[CI] 1--0}, and \mbox{[CI] 2--1} in \mbox{NGC~3603}.  We present fully sampled 200\arcsec$\times$200\arcsec maps integrated over the full velocity range as well as velocity channel maps. The observed field includes the central OB cluster, as well as the two adjacent molecular clumps MM1 and MM2, hosting two pillars observed by the HST and similar in appearance to the famous pillar structures in the M16 nebula. 
We selected two cuts from the OB cluster position to the peak emission in MM1 and MM2 for detailed analysis. The spectra along the two cuts show a rich kinematic structure, especially towards MM2. The observed maps show a strong correlation between the spatial distribution of C and CO. This implies either a face-on configuration of the clouds, i.e., both clouds being not in the same plane as the OB cluster and thus being illuminated face-on, or a more complex configuration, for instance a composition of many small clumps. The overall good C-CO correlation may indicate an unresolved, clumpy structure, but the increase of $\mathrm{N}_\mathrm{C}$ towards a cloud edge could hint towards relatively well shielded inner parts. 

Using escape probability model calculations we derived temperatures, densities, and column densities for both clouds from the observed line ratios. The temperatures are 43~K and 47~K for the peak positions of MM1 and MM2, respectively. We find gas densities of $n=10^{3}-10^4$~cm$^{-3}$ in MM1 and MM2.
% We find column densities of atomic carbon of 6 and 12$\times10^{17}$~cm$^{-2}$ in MM1 and MM2 and
From the best fit CO column densities we derive total H$_2$ column densities (filling corrected) of $0.9\times10^{22}$~cm$^{-2}$ and $2.5\times10^{22}$~cm$^{-2}$ for MM1 and MM2 and a ratio of  N$_\mathrm{C}$/N$_\mathrm{H_2}$ of 1-5$\times10^{-5}$.

The cluster is strongly interacting with the ambient molecular cloud and governing its structure and physical conditions. The stability analysis shows the existence of gravitationally bound gas which should lead to star formation. Embedded IR sources have already been observed in the outskirts of the molecular cloud and support our conclusions.  

We used HIRES/IRAS far-infrared data to narrow down the value of the FUV field at the positions of the two molecular clouds. Consistent with estimates from spectral type approximations and with [CII] observations by ISO we find $\chi\approx3-6\times 10^3$ in units of the Draine field.

However, many issues remain unresolved. For example, the analysis results are not conclusive regarding the clumpiness of the gas. NGC~3603 will be observed within the Herschel Guaranteed Time Key project {\it Warm and Dense ISM - WADI}. These data will allow us to perform a much more detailed study of the local gas conditions. Cooling lines like \OI\ and \CII\ as well as high-J CO lines can be used to study the clumpiness and the energy balance of the gas.

% rephrase
%Further observations with APEX and Herschel will provide the data to further study the physical and chemical conditions in NGC~3603 within the framework of PDR models. 

\begin{acknowledgements} %%%%%%%%%%%%%%%%%%%%%%%%%%%%%%%%%%%%%%%%%%%%%%%%  
  
  We made use of the
  NASA/IPAC/IRAS/HiRES data reduction facilities.  Data reduction of
  the spectral line data was done with the {\tt gildas} software
  package supported at IRAM (see {\tt
    http://www.iram.fr/IRAMFR/GILDAS}).

This work is financially supported in part by a Grant-in-
Aid for Scientific Research from the Ministry of Education,
Culture, Sports, Science and Technology of Japan (No.
15071203) and from JSPS (No. 14102003 and No. 18684003),
and by the JSPS core-to-core program (No. 17004). This work
is also financially supported in part by the 
German \emph{Deut\-sche For\-schungs\-ge\-mein\-schaft, DFG\/} 
 grants SFB494 and Os~177/1--1.
%
%of the Deutsche Forschungsgemeinschaft, the Ministerium
%f\"ur Innovation, Wissenschaft, Forschung und Technologie des
%Landes Nordrhein-Westfalen and through special grants of the
%Universit\"at Bonn and 
%Universit\"at zu K\"oln.

\end{acknowledgements}   
 
\bibliographystyle{aa} %%%%%%%%%%%%%%%%%%%%%%%%%%%%%%%%%%%%%%%%%%%%%%%%%%   
\bibliography{14765} % carina.bib
%\Online
\appendix
\section{Velocity Channel maps}
Here we present the velocity channel maps of all observed transitions. Each map
is shown at its original spatial resolution, i.e. 38'' for \mbox{CO 4--3}
and \mbox{\ion{C}{I} 1--0} as well as 26.5'' for \mbox{CO 7--6}
and \mbox{\ion{C}{I} 2--1}. 
\begin{figure*}[h]   
  \centering
  \includegraphics[angle=00,width=17cm]{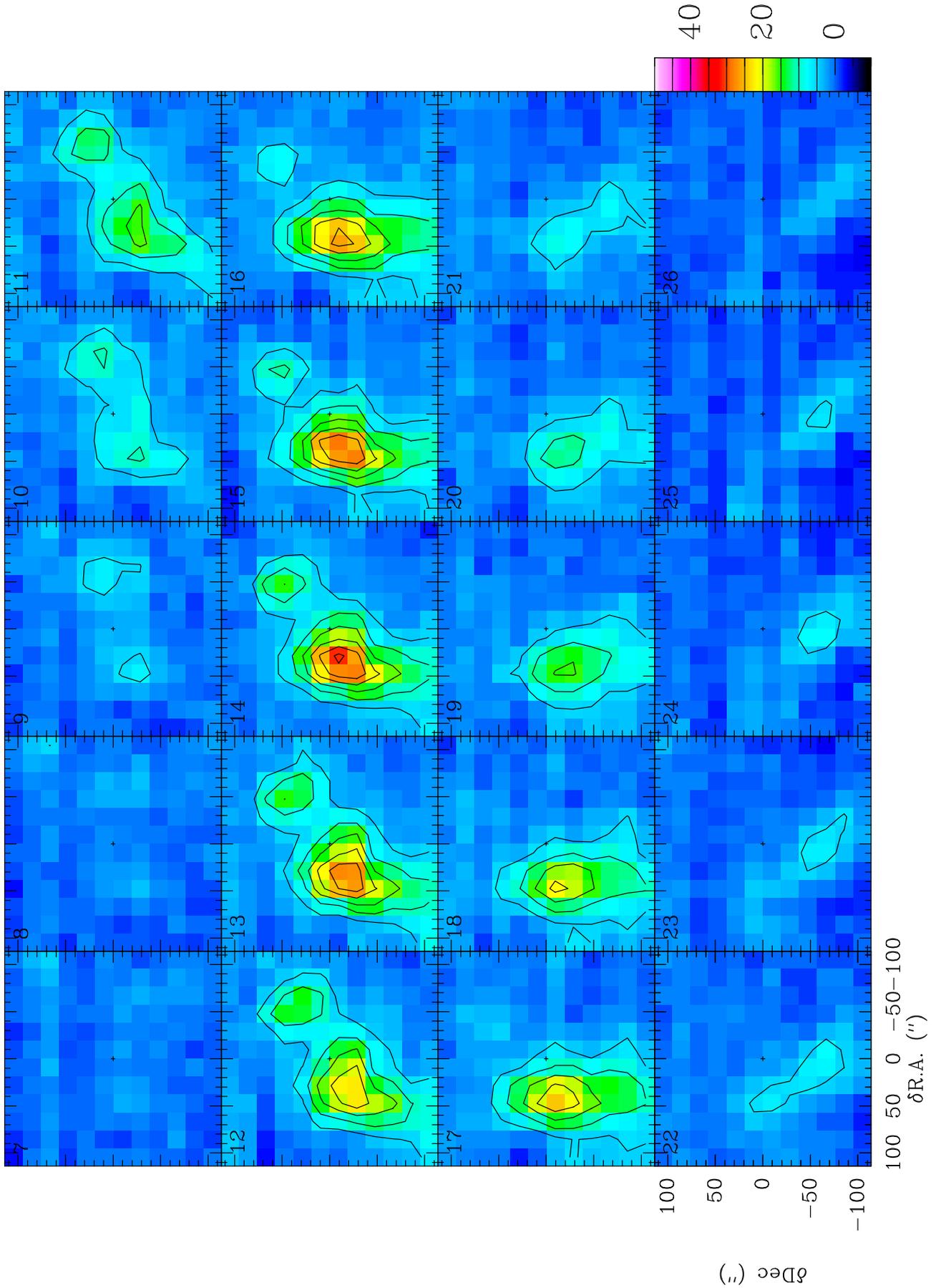}
\caption{Velocity structure of the observed field: \mbox{CO 4--3} intensity in velocity
channels of 1~km~s$^{-1}$ width.  Contours range between 5 and
45~K in steps of 5~K.  The position (0,0) corresponds to R.A.(J2000.0)=$\rasf{11}{15}{08}{85}$, 
Dec.(J2000.0)=$ \decasf{-61}{16}{50}{0}$. }
\label{co43-channels-apendix}   
\end{figure*}   
\begin{figure*}[h]   
  \centering
  \includegraphics[angle=00,width=17cm]{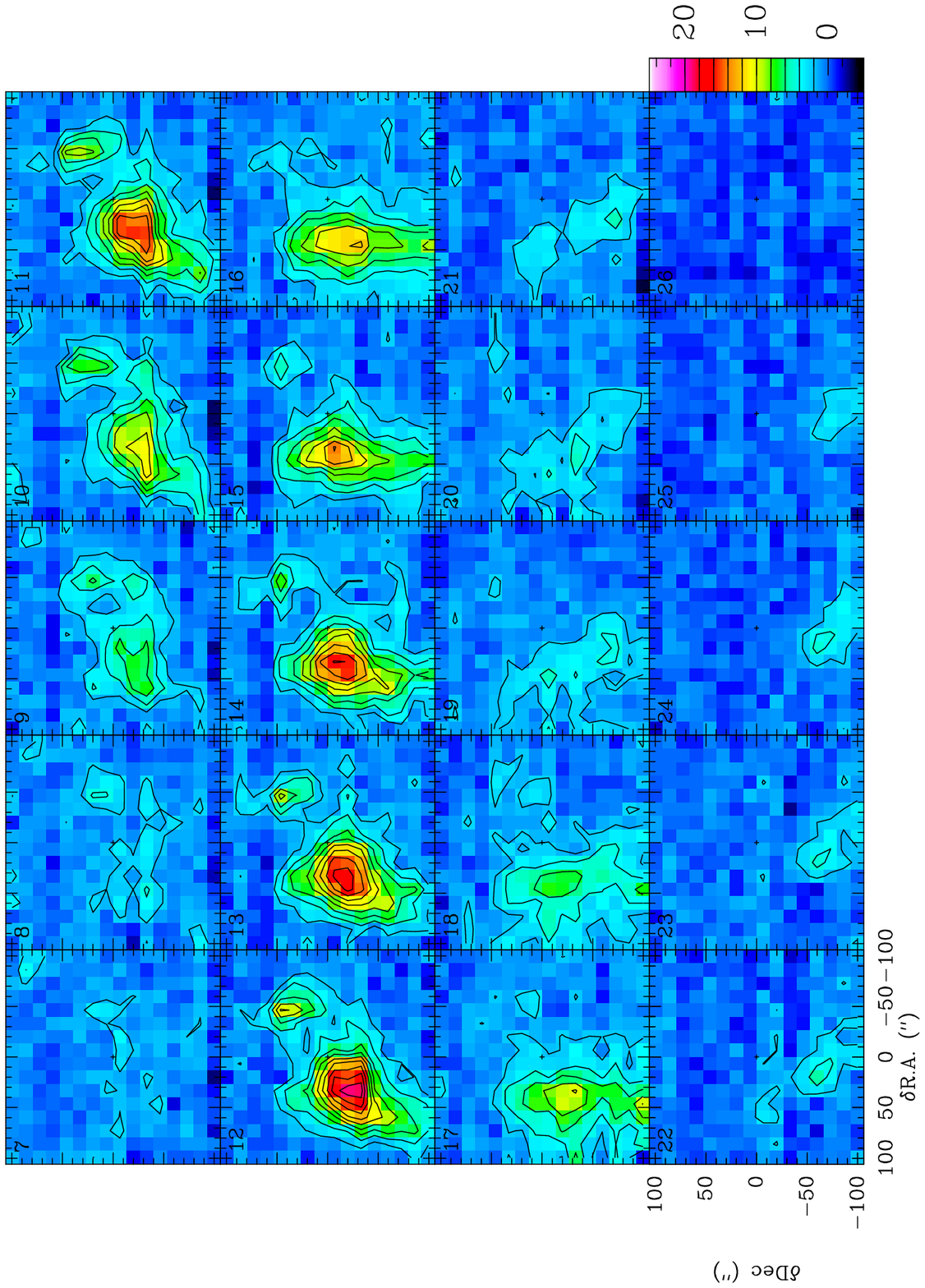}
\caption{Velocity structure of the observed field: \mbox{CO 7--6} intensity in velocity
channels of 1~km~s$^{-1}$ width.  Contours range between 2 and
20~K in steps of 2~K.  The position (0,0) corresponds to R.A.(J2000.0)=$\rasf{11}{15}{08}{85}$, 
Dec.(J2000.0)=$ \decasf{-61}{16}{50}{0}$.}
\label{co76-channels-apendix}   
\end{figure*} 
\begin{figure*}[h]   
  \centering
  \includegraphics[angle=00,width=17cm]{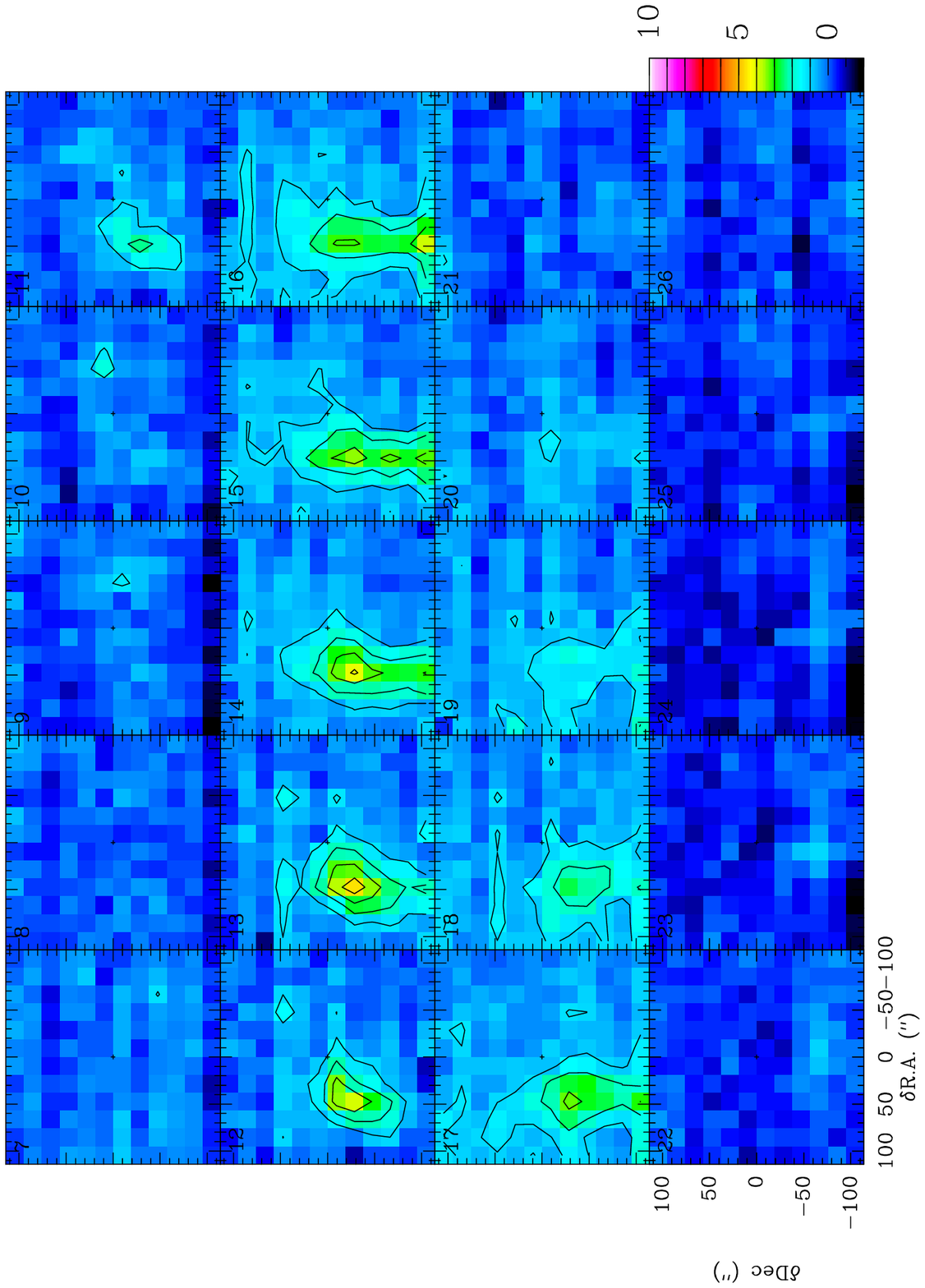}
\caption{Velocity structure of the observed field: \mbox{[CI] 1--0} intensity in velocity
channels of 1~km~s$^{-1}$ width.  Contours range between 1 and
10~K in steps of 1~K.  The position (0,0) corresponds to R.A.(J2000.0)=$\rasf{11}{15}{08}{85}$, 
Dec.(J2000.0)=$ \decasf{-61}{16}{50}{0}$.}
\label{ci10-channels-apendix}   
\end{figure*} 
\begin{figure*}[h]   
  \centering
  \includegraphics[angle=00,width=17cm]{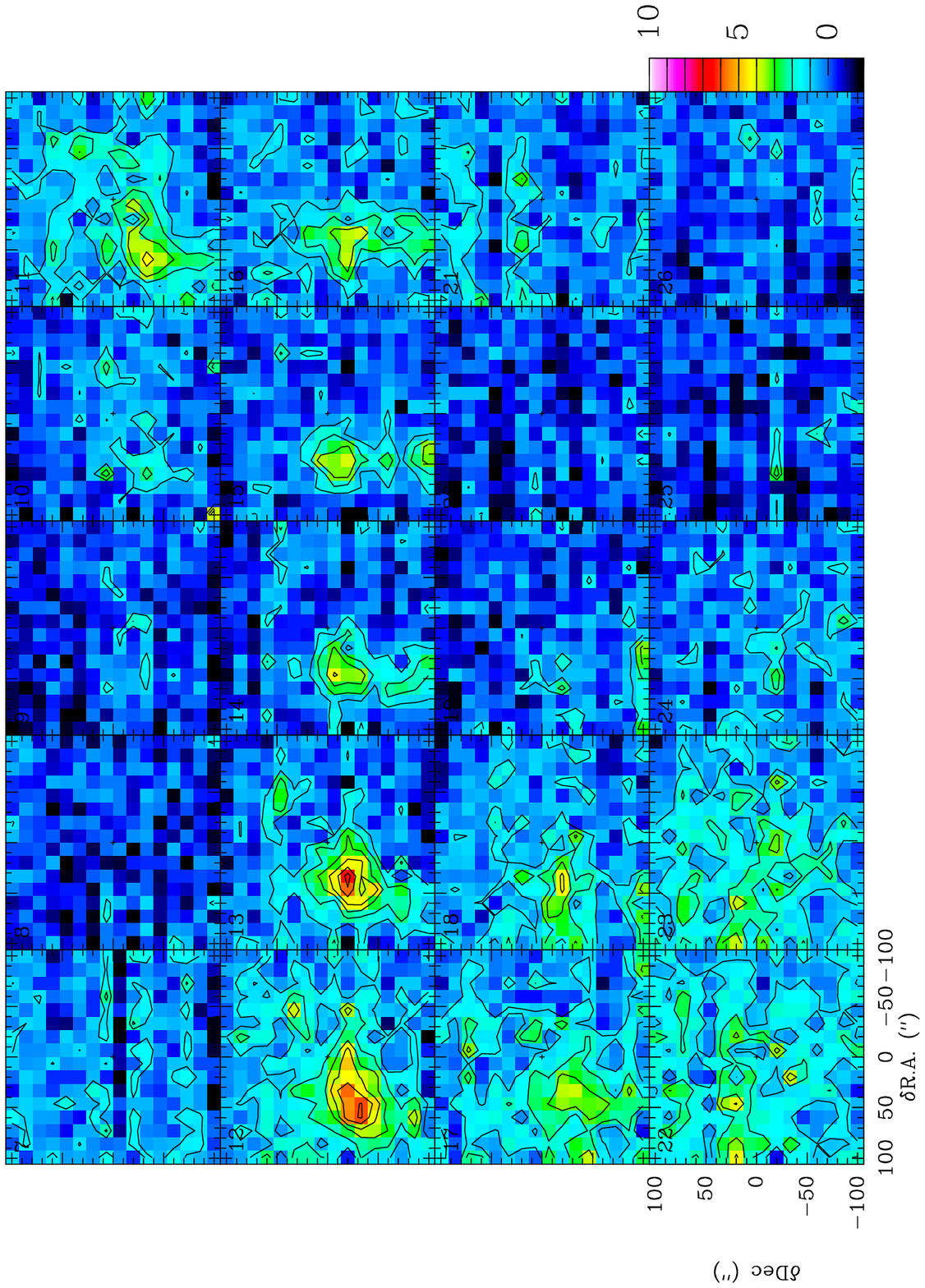}
\caption{Velocity structure of the observed field: \mbox{[CI] 2--1} intensity in velocity
channels of 1~km~s$^{-1}$ width.  Contours range between 1 and
10~K in steps of 1~K.  The position (0,0) corresponds to R.A.(J2000.0)=$\rasf{11}{15}{08}{85}$, 
Dec.(J2000.0)=$ \decasf{-61}{16}{50}{0}$.}
\label{ci21-channels-apendix}   
\end{figure*} 

\end{document}